\documentclass[a4paper]{jpconf}
\usepackage{graphicx}
\usepackage{lscape}
\usepackage{rotating}
\usepackage{colordvi}
\usepackage{color}
\usepackage[english]{babel}
\usepackage{amsmath}
\usepackage{amssymb}
\usepackage{amsfonts}
\usepackage{epsfig}
\usepackage{subfig}

\begin{document}
\title{Probing $f(R)$ gravity with PLANCK data on cluster pressure profiles }

\author{I. De Martino$^{1,4}$, M. De Laurentis$^{2,3,4}$, 
F. Atrio-Barandela$^{1}$, S. Capozziello$^{3,4}$}

\address{$^{1}$ F\'{\i}sica Te\'orica, Universidad de Salamanca, 37008 Salamanca, Spain; \\ 
$^{2}$Department of Theoretical Physics, Tomsk State Pedagogical University (TSPU), pr. Komsomolsky, 75, Tomsk, 634041, Russia;\\
$^{3}$ Dipartimento di Fisica, Universit\`a di Napoli "Federico II";\\
$^{4}$INFN sez. di Napoli Compl. Univ. di Monte S. Angelo, Edificio G,  
Via Cinthia, I-80126 - Napoli, Italy.}

\ead{ivan.demartino@usal.es}

\begin{abstract}
Analytical $f(R)$-gravity models introduce Yukawa-like corrections to the Newtonian
potential in the weak field limit. These models can explain the dynamics of galaxies and
cluster of galaxies without requiring dark matter. To test the model, 
we have computed the pressure profile of 579 X-ray galaxy clusters assuming
the  gas is in hydrostatic equilibrium within the potential well of the modified 
gravitational potential. We have compared those profiles
with the ones measured in the foreground cleaned SMICA released by
the Planck Collaboration. Our results show that Extended Theories 
of Gravity explain the dynamics of self-gravitating systems at
cluster scales and represent an alternative to dark matter haloes.
\end{abstract}

\section{Introduction}
The Cosmic Microwave Background (CMB) temperature anisotropies measured by the
Wilkinson Microwave Anisotropy Probe (WMAP) \cite{wmap9} and the Planck satellites
\cite{PLANCKXV2013, PLANCKXVI2013, PLANCKXX2013, PLANCKXXI2013} as well as luminosity
distances measured of the Supernovae Type Ia (SNeIa) \cite{suzuki},
strongly favor the concordance $\Lambda$CDM model, where the Universe
is dominated by two unknown energy densities: Dark Energy (DE) and Dark Matter (DM).
As an alternative, analytical $f(R)$-models modify the description of gravity
instead of requiring extra matter components. In these extensions
of General Relativity (GR), the Hilbert-Einstein action 
is replaced with a more  general function of the Ricci scalar $R$
\cite{PRnostro}.  These models have been found to introduce negligible 
corrections at the scale of the Solar systems \cite{berry} so they verify
the constraints imposed by the classical test of GR, and the 
perturbations in self-gravitating systems behave similarly as in
GR \cite{jeans}. Nevertheless, this  modified
gravity introduces a Yukawa-like correction to the gravitational potential
in the Post-Newtonian limit \cite{PRnostro, annalen}. This modified potential
has been used to explain the dynamics of galaxies 
without requiring DM \cite{cardone, napolitano}.

Since, the exact functional form of the Lagrangian is unknown,
it is important  to test potential models using all the available data
on self-gravitating systems. 
Galaxy clusters are reservoirs of hot gas and when CMB photons
cross the cluster potential well, they are scattered off by the free electron
of the intracluster medium. The effect, known as Sunyaev-Zeldovich (SZ) \cite{tsz,ksz},
induces secondary anisotropies that have been extensively measured  
\cite{PLANCKV2012, PLANCKX2012, PLANCKXX2013, PLANCKXXIX2013}.  
Pressure profiles of cluster of galaxies based on the Navarro-Frenk-White 
(NFW, \cite{NFW1997}) DM density profile have been found to be 
in agreement with SZ \cite{atrio2008} and X-ray observations \cite{arnaud2010}. 
We have constructed the pressure profiles of clusters of
galaxies assuming the gas is in hydrostatic equilibrium within
a Newtonian potential with a Yukawa correction and compared them with
Planck data. 
This article is organized as follows: in Sec.~2, we describe
the weak  field limit of $f(R)$ gravity; in Sec.~3, we summarize the 
pressure profiles most commonly used in the literature
and compare them with the pressure profile for $f(R)$ 
models; in Sec.~4  we describe the data used in our analysis; in
Sec.~5 we present our results and in Sec.~6 we will
summarize our main conclusions.  

\section{Yukawa corrections to the Newtonian potential
in $f(R)$-gravity} 

In Extended Theories of Gravity (ETGs), the field equations are obtained
by varying the action 
\begin{equation}\label{action}
{\cal A}=\frac{c^4}{16\pi G}\int d^4x \sqrt{-g}f(R) + {\cal L}_{m}\, ,
\end{equation}
resulting in \footnote{${\displaystyle f'(R)={df(R)}/{dR}}$ is the first derivative with 
respect to the Ricci scalar, $\Box_g={{}_{;\sigma}}^{;\sigma}$ 
is the d'Alembertian with covariant derivatives, 
$\displaystyle T_{\mu\nu}=-2(-g)^{-1/2}{\delta(\sqrt{-g}\mathcal{L}_m)/\delta g^{\mu\nu}}$ 
is the matter energy-momentum tensor, $T$ its trace, $g$ the determinant of
the metric tensor $g_{\mu\nu}$. Greek indices run from $0$ to $3$.}
\begin{equation*}\label{eq:fe1}
f'(R)R_{\mu\nu}-\frac{f(R)}{2}\,g_{\mu\nu}-
f'(R)_{;\mu\nu}+g_{\mu\nu}\Box_g f'(R)\,=\,8\pi G T_{\mu\nu}\,.
\end{equation*}

To construct spherically symmetric solutions, we write the metric element as
\begin{equation}
 ds^2=g_{tt}c^2dt^2 - g_{rr}dr^2 - r^2d\Omega,
\end{equation}
where $d\Omega$ is the solid angle. We restrict our study to
$f(R)$-Lagrangians that are  expandable in Taylor series around a fixed point $R_0$ 
\begin{equation}\label{eq:sertay}
f(R)=\sum_{n}\frac{f^n(R_0)}{n!}(R-R_0)^n\simeq
f_0+f'_0R+\frac{f''_0}{2}R^2+...\,.
\end{equation}
In the Post-Newtonian limit of these models
the gravitational potential can be written as \cite{PRnostro}
\begin{equation}
\Phi_{grav}(r)=\frac{\Phi_N(r)}{(1+\delta)}\left(1+\delta e^{-\frac{r}{L}}\right).
\label{eq:pot}
\end{equation}
In this expression, $\delta$ represents the deviation from GR at zero order and the  
Newtonian limit of GR is recovered at $\delta\rightarrow0$, irrespective
of the scale parameter $L$. The latter is the 
scale length of the self-gravitating object \cite{PRnostro, annalen} and
its effects are negligible at Solar System scales where GR, 
is restored \cite{annalen}.  In terms of the coefficients
in the Taylor expansion they are given by $\delta=1-f'_0=1$ 
and $L=[-{f'_0}/{(6f''_0)}]^{1/2}$. In this context, the dynamical
effects of DM are now given by the modifications of the Newtonian potential.

\section{Cluster pressure profiles in $f(R)$ gravity}

SZ effect induces two different type of secondary temperature anisotropies on the CMB: a
thermal  contribution due to the motion of the electrons within the cluster
potential well and a kinematic one (KSZ) due to the peculiar motion of the cluster. 
It is given by
\begin{equation}
\frac{\Delta T}{T_0}=g(\nu)\frac{k_B\sigma_T}{m_ec^2}\int n_eT_e dl,
\label{eq:sz}
\end{equation}
where $T_e$ is the electron temperature, $n_e$ the electron density and the 
integration is carried out along the line of sight $l$. In Eq.~(\ref{eq:sz}) 
$k_B$ is the Boltzmann constant, $m_ec^2$ the electron annihilation 
temperature, $c$ the speed of light, $\nu$ the frequency 
of observation, $\sigma_T$ the Thomson cross section and
$T_0$ the mean temperature of the CMB. Finally, $g(\nu)=x\coth(x/2)-4$
is the frequency dependence of the TSZ effect, with the reduced frequency 
$x=h\nu/KT_0$. For individual clusters, only the TSZ anisotropy has been measured.

Eq.~(\ref{eq:sz}) shows that in order to compute the TSZ anisotropies 
we need to estimate the cluster pressure profile, $n_eT_e$.
Several  cluster profiles, based on  X-ray data and numerical simulations, 
have appeared in the literature: 

$\bullet$ The isothermal $\beta$-model 
\cite{cavaliere1976, cavaliere1978} specifies 
the electron density: $n_e(r)=n_{e,0}[1+(r/r_c)^2]^{-3\beta/2}$.
From X-ray surface brightness data it has been found that $\beta=0.6-0.8$ \cite{jones1984}. 

$\bullet$ The universal profile \cite{arnaud2010} is a phenomenological 
model based on the NFW DM-profile
\begin{equation}
p(x) \equiv \frac{P_0}{(c_{500}x)^{\gamma_a} 
[1+(c_{500}x)^{\alpha_a}]^{(\beta_a-\gamma_a)/\alpha_a}},
\label{eq:universal_profile}
\end{equation}
where $x$ is the radial distance in units of $r_{500}$,
the radius where the average density is 500 times the critical density,
and $c_{500}$ is the concentration parameter at $r_{500}$. Different
groups have fit the model parameters $[c_{500},\alpha_a, \beta_a, \gamma_a, P_0]$ 
to X-ray or CMB data; their best fit values are given in Table~\ref{table1}.

\begin{figure}[!ht]
\centering 
\includegraphics[width=0.5\columnwidth]{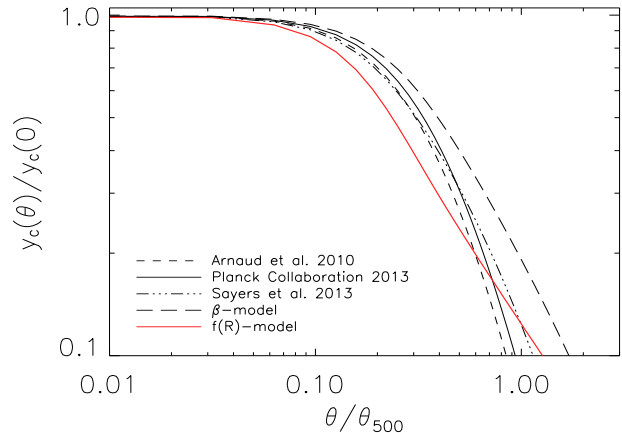}\\
\caption{
Pressure profiles integrated along the line of sight
for the Coma cluster ($z=0.023$). We plot three GNFW profiles
(dashed, solid and dash-dotted lines), one $\beta=2/3$ model (long
dashed line) and a $f(R)$ model (red solid line). The parameters of each model
are given in Table~\ref{table1}.
}
\label{Fig1}
\end{figure}

\begin{table}
\caption{Parameters of the GNFW corresponding to 
\cite{arnaud2010,PLANCKV2012,Sayers2012}, $\beta$ for Coma cluster
and $f(R)$ is the best fit model to Planck data.}\label{table1}
\centering
\small{\begin{tabular}{|lccccc|}
\hline
\emph{Model} & $c_{500}$ & $\alpha_a$ & $\beta_a$ & $\gamma_a$ & $P_0$ \\ 
\hline
\emph{Arnaud} & 1.177 & 1.051 & 5.4905 &  0.3081 & $8.403h_{70}^{3/2}$ \\ 
\emph{Planck} & 1.81 & 1.33 & 4.13 & 0.31 & 6.41 \\
\emph{Sayers} & 1.18 & 0.86 & 3.67 & 0.67  &  4.29 \\
\br
\br
\emph{$\beta$} & $\beta$ & $n_{c,0}/m^{-3}$ & $r_c$/Mpc & $T_e$/keV    & \\
\hline
                     & 2/3     &   3860.   & 0.25  &  6.48         & \\
\br
\br
\emph{$f(R)$} & $\delta$ & L/Mpc & $\gamma$ & &  \\
\hline
                    &   -0.98  &  0.1    &  1.2     & &  \\
\hline
\end{tabular}}
\end{table}
To compute in $f(R)$ gravity the pressure profile $n_eT_e$ of Eq.~(\ref{eq:sz}) 
we assume  that the gas is in hydrostatic equilibrium within the 
modified potential well of the cluster (without the DM contribution)
       \begin{equation}\label{eq:HE}
        \frac{dP(r)}{dr}=-\rho(r)\frac{d\Phi_{grav}(r)}{dr},
       \end{equation}
and the physical state of the gas is described by a polytropic equation of state 
       \begin{equation}\label{eq:PES}
        P(r)\propto\rho^\gamma(r).
       \end{equation}
Eqs.~\eqref{eq:HE} and \eqref{eq:PES}, together with the continuity equation
        \begin{equation}\label{eq:EMC}
        \frac{dM(r)}{dr}= 4\pi\rho(r),
       \end{equation}
and the cluster gravitational potential given by eq.~(\ref{eq:pot})
form a close system that can be solved 
numerically to obtain the pressure profiles of any given cluster as a function
of two gravitational parameters $(\delta, L)$ and the polytropic index $\gamma$.
Fig.~\ref{Fig1} we represent the different profiles integrated along the line
of sight with the parameters given in Table~\ref{table1}. The profiles are
particularized for the parameters of the Coma cluster.

\section{Data.}

We use Planck data and a X-ray selected cluster catalog 
to constrain the pressure profiles of clusters of galaxies.

$\bullet$ Our cluster catalog is fully described in \cite{catalog} and 
lists  position, redshift, X-ray flux, X-ray luminosity.
The X-ray electron temperature is derived from the $L_X-T_X$ relation of \cite{white}. 
A total of 579 clusters are located outside the minimal Planck mask;
the region where the CMB data is less affected by galactic foreground residuals.

$\bullet$ The Planck Collaboration released in
March 2013 nine maps of the CMB temperature anisotropies
spanning a frequency range from 30 to 845 GHz.  Since foreground
cleaned maps at all frequencies were not made available, we 
performed our analysis on a foreground cleaned map known
as SMICA. This map has been constructed using a component
separation method by combining the data at all frequencies
\cite{PLANCKXII2013}. The SMICA map has $5'$ resolution.

For all except the brightest clusters, the underlying CMB anisotropies
have large amplitude than the TSZ contributions. Then, to compare
the expected profile with the data, we stack the signal of all the 
clusters in our sample. The anisotropy is averaged in 
rings of width $\theta_{500}/2$, where $\theta_{500}$ is 
the angular scale subtended by the $r_{500}$. For each data point 
an error bar is obtained by evaluating 1,000 times the average profiles 
at 579 random positions in the SMICA map. To avoid overlapping real
and simulated clusters we removed a disc of $80'$ around
each of the clusters in our sample.

Using eqs.~\eqref{eq:pot}, \eqref{eq:HE}, 
\eqref{eq:PES} and \eqref{eq:EMC}, we constructed the pressure
profile of all clusters in the data as a function
of three parameters: $(\delta, L,\gamma)$. We integrated the profiles
along the line of sight and convolved them with a Gaussian beam
with the same resolution of the SMICA map. 
We considered two parameterizations. In (A) $L= \zeta r_{500}$ 
differs from cluster to cluster but scales linearly with 
$r_{500}$; in this parametrization we assume that $\zeta=[0.1,4]$.
In (B) $L$ has the same value for all the clusters; it varies in the range 
$L=[0.1,20]$Mpc, from the scale of a typical core radius to the mean cluster 
separation.  Since if $\delta<-1$ the potential is repulsive and if 
$\delta=-1$ the potential diverges, we took $\delta=[-0.99,1.0]$.
Finally, the polytropic index $\gamma$  was set to vary
in the range $\gamma=[1.0,1.6]$, that corresponds to an isothermal 
and adiabatic monoatomic gas, respectively. We took 30 equally spaced steps in
all intervals.

\section{Results and discussion}
We computed the likelihood function $\log{\cal L}=-\chi^2/2$ as
\begin{equation}
\chi^2 ({ p})=\Sigma_{i,j=0}^{N} 
(y({ p}, x_i)-d(x_i))C_{ij}^{-1}
(y({ p}, x_j)-d(x_j))
\label{eq:chi}
\end{equation}
where $N=7$ is the number of data points, and
${ p}=(\delta,L,\gamma)$. In eq.~(\ref{eq:chi}), $d(x_i)$ is the data
and $C_{i,j}$ is the correlation function between the average
temperature anisotropy on discs and rings in bins of size $\theta_{500}$. The process
was computed on 579 random  positions outside the cluster locations  and repeated
1,000 times. 

\begin{figure*}
\centering 
\includegraphics[width=0.95\columnwidth]{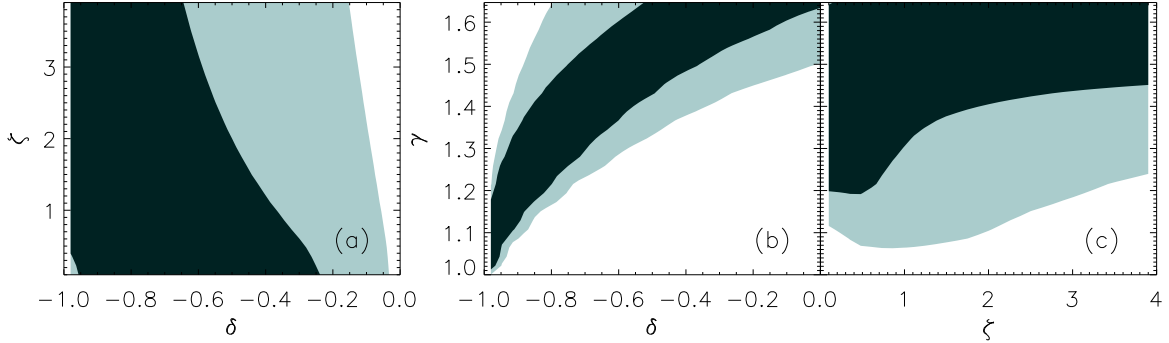}\\
\caption{Confidence contours for pairs of parameters of
Model A. Contours are at the 68\% and 95\% confidence level. 
}\label{Fig4}
\end{figure*}

\begin{figure*}
\centering 
\includegraphics[width=0.95\columnwidth]{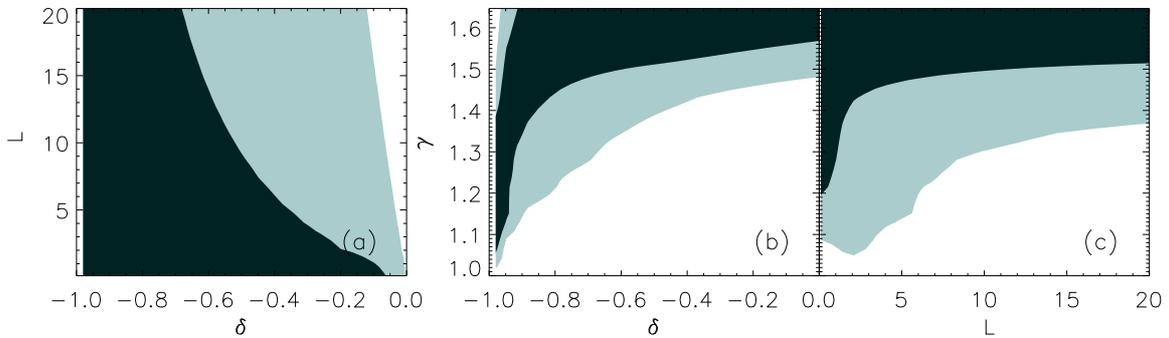}\\
\caption{Same as in  Fig. \ref{Fig4} for Model B.} 
\label{Fig5}
\end{figure*}

In Figs.~\ref{Fig4} and \ref{Fig5}, we computed the 2D contours at 
the 68\% and 95\% confidence levels of the marginalized likelihoods of pairs 
of parameters of Model A and Model B, respectively. Since the contours are 
not closed, we can only quote the following upper limits at the same level of confidence:
$\delta<-0.46,-0.10$, $\zeta < 2.5,3.7$ and $\gamma>1.35,1.12$ for Model A
and $\delta<-0.43,-0.08$, $L<12,19$ Mpc and $\gamma>1.45,1.2$ for Model B. 
Notice that $L$ is similar in both A and B and 
$\gamma$ dominated by the physical boundary imposed. 

Although model parameters are weakly constrained by the data,
the value $\delta=0$ is excluded at more than 95\% confidence level.  
Since $\delta\simeq 0$  corresponds to the standard Newtonian potential 
without DM, our result indicates that a self-gravitating gas,
without DM or modified gravity, does not fit the data.

\section{Conclusions}
We have constrained the analytical $f(R)$ model of gravity by comparing the
pressure profile of clusters of galaxies with CMB data.
We have demonstrated that these models can accurately fit
cluster profiles without a potential well dominated by DM. 
In this context, $f(R)$ gravity can be considered a viable alternative
to DM. Our results also showed that a profile made of gas, without
a modified potential or without significant DM contributions could not
fit the observed profiles.

Our constraints on model parameters are rather weak since we only used
the data on the cluster profile but we did not use the frequency dependence
of the TSZ effect, the reason being that the Planck Collaboration has
not released foreground cleaned maps at each frequency. Then, our results
could be substantially improved when including this frequency information,
opening the possibility of constraining - ruling out the modified Newtonian
potential at this scale.

\end{document}